\documentstyle[12pt]{article}
\begin{document}
\thispagestyle{empty}
\newcommand{\be}{\begin{equation}}
\newcommand{\ee}{\end{equation}}
\newcommand{\sect}[1]{\setcounter{equation}{0}\section{#1}}
\newcommand{\vs}[1]{\rule[- #1 mm]{0mm}{#1 mm}}
\newcommand{\hs}[1]{\hspace{#1mm}}
\newcommand{\mb}[1]{\hs{5}\mbox{#1}\hs{5}}
\newcommand{\bea}{\begin{eqnarray}}
\newcommand{\eea}{\end{eqnarray}}
\newcommand{\wt}[1]{\widetilde{#1}}
\newcommand{\ux}[1]{\underline{#1}}
\newcommand{\ov}[1]{\overline{#1}}
\newcommand{\sm}[2]{\frac{\mbox{\footnotesize #1}\vs{-2}}
           {\vs{-2}\mbox{\footnotesize #2}}}
\newcommand{\prt}{\partial}
\newcommand{\eps}{\epsilon}\newcommand{\p}[1]{(\ref{#1})}
\newcommand{\R}{\mbox{\rule{0.2mm}{2.8mm}\hspace{-1.5mm} R}}
\newcommand{\Z}{Z\hspace{-2mm}Z}
\newcommand{\cd}{{\cal D}}
\newcommand{\cg}{{\cal G}}
\newcommand{\ck}{{\cal K}}
\newcommand{\cw}{{\cal W}}
\newcommand{\vj}{\vec{J}}
\newcommand{\vl}{\vec{\lambda}}
\newcommand{\vz}{\vec{\sigma}}
\newcommand{\vt}{\vec{\tau}}
\newcommand{\poiss}{\stackrel{\otimes}{,}}
\newcommand{\tx}{\theta_{12}}
\newcommand{\tb}{\overline{\theta}_{12}}
\newcommand{\zw}{{1\over z_{12}}}
\newcommand{\sqp}{{(1 + i\sqrt{3})\over 2}}
\newcommand{\sqm}{{(1 - i\sqrt{3})\over 2}}
\newcommand{\NP}[1]{Nucl.\ Phys.\ {\bf #1}}
\newcommand{\PLB}[1]{Phys.\ Lett.\ {B \bf #1}}
\newcommand{\PLA}[1]{Phys.\ Lett.\ {A \bf #1}}
\newcommand{\NC}[1]{Nuovo Cimento {\bf #1}}
\newcommand{\CMP}[1]{Commun.\ Math.\ Phys.\ {\bf #1}}
\newcommand{\PR}[1]{Phys.\ Rev.\ {\bf #1}}
\newcommand{\PRL}[1]{Phys.\ Rev.\ Lett.\ {\bf #1}}
\newcommand{\MPL}[1]{Mod.\ Phys.\ Lett.\ {\bf #1}}
\newcommand{\BLMS}[1]{Bull.\ London Math.\ Soc.\ {\bf #1}}
\newcommand{\IJMP}[1]{Int.\ J.\ Mod.\ Phys.\ {\bf #1}}
\newcommand{\JMP}[1]{Jour.\ Math.\ Phys.\ {\bf #1}}
\newcommand{\LMP}[1]{Lett.\ Math.\ Phys.\ {\bf #1}}
\renewcommand{\thefootnote}{\fnsymbol{footnote}}
\newpage
\setcounter{page}{0}
\pagestyle{empty}
\vs{12}
\begin{center}
{\LARGE {\bf Real Structures in Clifford Algebras and}}\\
{\LARGE {\bf Majorana Conditions in Any Space-time}}\\
[0.8cm]

\vs{10} {\large M.A. De Andrade$^{(a,\ast )}$ and F.
Toppan$^{(b,\ast )}$} ~\\ \quad \\ {\em {~$~^{(a)}$} UCP, FT, Rua
Bar{\~{a}}o do Amazonas, 124, cep 25685-070, Petr\'{o}polis (RJ),
Brazil}\\ ~\quad\\ {\em ~$~^{(b)}$ UFES, CCE Depto de
F{\'{\i}}sica, Goiabeiras cep 29060-900, Vit\'oria (ES), Brazil}\\
~\quad \\ {\em {~$^{(\ast ) }$} CBPF, DCP, Rua Dr. Xavier Sigaud
150, cep 22290-180 Rio de Janeiro (RJ), Brazil}

\end{center}
\vs{6}

\centerline{ {\bf Abstract}}

\vs{6}

Clifford algebras and Majorana conditions are analyzed in any
spacetime. An index labeling inequivalent $\Gamma$-structures up
to orthogonal conjugations is introduced.\par
Inequivalent charge-operators in even-dimensions,
invariant under Wick rotations, are considered. The hermiticity
condition on free-spinors lagrangians is presented. The
constraints put by the Majorana condition on the free-spinors
dynamics are analyzed. Tables specifying which spacetimes admit
lagrangians with non-vanishing kinetic, massive or pseudomassive
terms (for both charge-operators in even dimensions) are given.
The admissible free lagrangians for free Majorana-Weyl spinors are
fully classified.

\vs{6} \vfill \rightline{April 1999} \rightline{CBPF-NF-013/99}
\rightline{hep-th/xxx} {\em E-Mail:\\ 1) marco@cat.cbpf.br\\ 2)
toppan@cat.cbpf.br}
\newpage
\pagestyle{plain}
\renewcommand{\thefootnote}{\arabic{footnote}}
\setcounter{footnote}{0}
\vs{8}

\section{Introduction.}

The theory of Clifford algebras is an old subject which has been
extensively investigated both in the mathematical and in the
physicists' literature. For obvious reasons physicists mainly
dealt with the theory of spinors in Minkowskian or Euclidean
spacetimes \cite{{Glio},{Sche}}. Nevertheless in some papers
[3--6] spinors in pseudoeuclidean spacetimes with arbitrary
signature $(t,s)$, $D=t+s$ being the dimensionality of the
spacetime, have been analyzed.
In particular \cite{Kugo} can be regarded as the reference work on
the subject since it presents a rather complete list of results in
this topic.\par The development of supergravities and superstring
theories which emphasize the Kaluza-Klein aspect of
compactification to lower dimensional space led investigating
properties of spinors (and supersymmetries) in arbitrary
dimensional spaces. However, apart some special papers as the ones
previously recalled, the great majority of works were still
devoted to standard-signature spacetimes. The question of
providing a physical interpretation for the extra-times somehow
masked the fact that from a strictly mathematical point of view
consistent superstring theories can be formulated in exotic
signatures (like e.g. $5+5$). This negative attitude towards
exotic signatures seems at present time changing and their
possible physical implications find increasing attentions (see
e.g. \cite{Bars}).  Various reasons are at the basis of this shift
of attitude. Some recent works \cite{Hull} for instance pointed
out the existence of dualities relating theories formulated in
different signatures. On the other hand the still-mysterious
$M$-theory suggests that we need investigating along all possible
directions. A rather formal argument can also be invoked, a
reasonable demand for any possible theory which could claim to be
a genuine ``theory of everything" is that the signature of the
spacetime should be determined by the properties of the spacetime
itself rather than imposed a priori. The Minkowskian signature
should therefore be selected after confrontation with the other
signatures.
\par
Motivated by the above considerations in this paper we analyze the
real Clifford structures and Majorana conditions in any signature
spacetime with arbitrary dimensions. Without loss of generality,
the analysis here presented
is based on a specific Weyl realization of Clifford algebras. The
technique employed allows to recover the results of \cite{Kugo} in
a considerably simplified manner. Besides that,
extra-informations, not presented in \cite{Kugo}, are gained. As
an example discussed in the text we mention the correct choice of
the charge operator which preserves the Majorana condition under a
Wick rotation to the Euclidean.
\par
A list of further topics here discussed, some of them we are not
aware to be found elsewhere, is the following. An index is
introduced to label and discriminate among classes of inequivalent
Clifford algebras up to orthogonal conjugations. Such index could
in principle be relevant to physical applications whenever some
kind of reality conditions are imposed on the fields.
\par
Moreover the compatibility of the Majorana condition with the free
massive equations of motion is thoroughly investigated. The
conditions upon free lagrangians for Majorana spinors in order to
be non-vanishing, hermitian and charge-conjugated are presented.
Explicit and easy-to-consult tables of spacetimes supporting
massive (or pseudomassive in the even-dimensional case) Majorana
spinors are provided. The list of results here presented is more
complete than the one given in reference \cite{DeAn}.\par

\par
The scheme the present work is as follows: the next section is
devoted to notations and preliminary results. In section $3$ the
specific Weyl representation employed is introduced. Even-dimensional
inequivalent
charge operators, invariant under Wick rotations, are constructed.
The index labeling $\Gamma$-structures up to orthogonal
conjugation is discussed in section $4$. The hermiticity condition
on free-spinors lagrangians is discussed in section $5$. Section
$6$ presents an exhaustive list of the constraints put by the
Majorana condition on the free-spinors dynamics, both at the level
of the equations of motion and of the action. Tables specifying
which spacetimes admit lagrangians with non-vanishing kinetic,
massive or pseudomassive terms (for both charge-operators in even
dimensions) are given. Finally in section $7$ the problem of
determining the admissible free lagrangians (kind of terms,
non-vanishing conditions, type of coefficients) for Majorana-Weyl
spinors in even-dimensional spacetimes is fully solved.

\section{Notations and preliminary results.}

Let $\eta^{\mu\nu}$ be the (pseudo)-euclidean metric associated to
an $M^{t,s}$ generalized Minkowski space-time with $t$
time-directions and $s$ space-directions. The space-time dimension
being $D=t+s$. In the following we will denote as time (space)
directions those which are related to the $+$ (and respectively
$-$) sign in $\eta^{\mu\nu}$.\par A $\Gamma$-structure associated
to the $M^{t,s}$ spacetime is a matrix-representation of the
Clifford algebra generators $\Gamma^\mu$ ($\mu =1,...,D$)
satisfying the anticommutation relations
\begin{eqnarray}
\Gamma^\mu\Gamma^\nu +\Gamma^\nu\Gamma^\mu &=& 2\eta^{\mu\nu}
\cdot {\bf 1}_{\Gamma} \label{1}
\end{eqnarray}
The representation is realized by $2^{[{D\over 2}]}\times
2^{[{D\over 2}]}$ matrices ($[{D\over 2}]$ denoting the integral
part of ${D\over 2}$) which can be further assumed to satisfy the
unitarity requirement
\begin{eqnarray}
{\Gamma^{\mu}}^\dag &=& {\Gamma^{\mu}}^{-1} \label{2}
\end{eqnarray}
A tracelessness condition holds
\begin{eqnarray}
tr \Gamma^\mu &=& 0 \label{3}
\end{eqnarray}
for any $\mu$.\par According to the fundamental Pauli theorem
\cite{Saku} the above matrix-representation is uniquely realized
up to unitary conjugation. \par Notice that choosing the ($+$)
sign in the r.h.s. of (\ref{1}) is a matter of convention. The
opposite choice (i.e. r.h.s. $\equiv -2\eta^{\mu\nu}$) is
admissible. However it is just sufficient to look at results and
tables obtained for an $(s,t)$-signature within the $+$ convention
since the results for the $-$ convention-case are immediately
recovered by interchanging $t$ and $s$: $t\leftrightarrow s$. It
should be therefore clear that the complete set of solutions for
$(s,t)$ spacetimes are recovered from the tables produced below
only after such ($t,s$)-``dualization" has been taken into
account.
\par The introduction of lagrangians and charge conjugations for
the spinor fields require the presence of three (only two of them
mutually independent) unitary matrices, denoted in the literature
as $A,B,C$, associated to each one of the three conjugations
(hermitian, complex-conjugation and transposition respectively)
acting on the $\Gamma^\mu$-matrices, according to
\begin{eqnarray}
A\Gamma^\mu A^\dag &=& (-1)^{t+1} {\Gamma^\mu}^\dag\nonumber\\
B\Gamma^\mu B^\dag &=&\eta{\Gamma^\mu}^\ast\nonumber\\
C\Gamma^{\mu}C^\dag &=& \eta (-1)^{t+1}{\Gamma^\mu}^T \label{4}
\end{eqnarray}
As discussed later $\eta$, as well as $\varepsilon$ introduced
below, is a sign ($\pm 1$) specifying the assignment of a
$\{\Gamma^\mu, A, B, C\}$ structure up to unitary transformations.
$\eta$ and $\varepsilon$ will be explicitly computed in the next
section. The introduction of $\eta$ as defined in (\ref{4})
corresponds to the standard convention in the literature.\par The
equation relating $A,B$ and $C$ can be expressed through
\begin{eqnarray}
C&=& B^T A \label{5}
\end{eqnarray}
with the transposed matrix $B^T$ satisfying
\begin{eqnarray}
B^T &=& \varepsilon B \label{6}
\end{eqnarray}

An useful form of restating the above equation is
\begin{eqnarray}
B^\ast B &=& \varepsilon \cdot {\bf 1} \label{7}
\end{eqnarray}

The $A$-matrix can be expressed through the position
\begin{eqnarray}
A &=& \prod_{i=1,...,t}\Gamma^i \label{8}
\end{eqnarray}
where the product (the order is irrelevant since $A,B,C$ can
always be determined up to an arbitrary phase) is restricted to
time-like $\Gamma$-matrices, i.e. those satisfying the
${\Gamma^i}^2 =+{\bf 1}$ equation (conversely the spacelike
$\Gamma$-matrices are those belonging to the complementary set
satisfying ${\Gamma^j}^2=-{\bf 1}$).\par The $A$-matrix allows
constructing in generic flat spacetimes $M^{t,s}$ the conjugated
$\overline{\psi}$ spinor as $\overline{\psi}=\psi^{\dag}A$, and
generalizes the $\Gamma^{0}$-matrix of the standard Minkowskian
spacetime.\par The matrix $C$ corresponds to the
charge-conjugation matrix, while $B$ is employed in introducing
the charge-conjugated spinors $\psi^{c}$ according to
\begin{eqnarray}
\psi^{c}&=& B^{\dag}\psi^{\ast} \label{9}
\end{eqnarray}
Quantum mechanical states are rays in a Hilbert space. A physical
spinorial state can be equally well described by a spinor
transformed via an unitary matrix $U$, $\psi \mapsto U\psi$.\par
It is easily proven that under such a unitary transformation
$\Gamma^{\mu}, A,B,C$ are mapped as follows
\begin{eqnarray}
\Gamma^{\mu} &\mapsto& U\Gamma^\mu U^\dag\nonumber\\ A&\mapsto&
UAU^{\dag}\nonumber\\ B&\mapsto& U^\ast B U^{\dag}\nonumber\\
C&\mapsto & U^\ast C U^\dag \label{10}
\end{eqnarray}
Notice that the unitary transformations acting upon $B,C$ {\em do
not} coincide with their unitary conjugations.\par If we introduce
the notion of a $\{\Gamma^{\mu}, A, B, C\}$-structure assignment
associated to a given spacetime $M^{t,s}$ and we look for
inequivalent classes of such assignments under the (\ref{10})
transformations, we easily realize that the $\eta, \varepsilon$
signs introduced above label inequivalent classes of assignments.
Indeed $\eta,\varepsilon$ can be equivalently introduced in a
unitary-invariant trace form as
\begin{eqnarray}
tr(B^\ast B) &=& \varepsilon \cdot tr {\bf 1}\nonumber\\
tr(B\Gamma^{\mu}B^\dag{\Gamma_\mu}^{\ast} )&=& \eta D\cdot tr{\bf
1} \label{11}
\end{eqnarray}
where the convention on the repeated indices is understood.\par
With a slight abuse of language we can say that $\eta,\varepsilon$
label inequivalent choices of charge-conjugations.\par In an
even-dimensional spacetime ($D=2n$) we can introduce a timelike
generalized $\Gamma^5$ matrix (i.e. the matrix generalizing the
one associated to the ordinary Minkowskian spacetime), through the
position
\begin{eqnarray}
\Gamma^5 &=& (-1)^{{s-t\over 4}}\prod_{\mu = 1, ..., D} \Gamma^\mu
\label{12}\end{eqnarray}

The sign is chosen in order to guarantee ${\Gamma^5}^2={\bf
1}$.\par Let us conclude this section by presenting some further
useful identities
\begin{eqnarray}
A^\dag&=& (-1)^{{t\over 2}(t-1)} A\nonumber\\ A^\ast &=&\eta^t B A
B^\dag\nonumber\\ A^T &=& \eta^t (-1)^{{t\over 2}(t-1)} CAC^\dag
\label{13}
\end{eqnarray}
and
\begin{eqnarray}
C^T &=& \varepsilon \eta^t (-1)^{{t\over 2}(t-1)} C \label{14}
\end{eqnarray}

\section{Clifford algebras and the Majorana condition.}

The allowed values for the signs $\eta,\varepsilon$ labelling
inequivalent $\{\Gamma^\mu, A,B,C\}$-structures associated to any
given spacetime have been computed in \cite{Kugo}. A very
efficient and much simpler method of computing $\eta,\varepsilon$
is at disposal by explicitly using a $\Gamma$-structure in a
given Weyl
representation. The choice of working within such Weyl representation
can always be done and, due to the fundamental property that
$\Gamma$-structures are all equivalent up to unitary conjugation
\cite{Saku}, by no means affects the generality of the results so
obtained. More than just reproducing previous results the
computation with the choice we made encodes further
information. Indeed we will show that our $C$
charge-conjugation operators are left unchanged under a Wick
rotation. When inequivalent charge-conjugation operators are
present the tables provided below inform which charge-conjugations
should be correctly chosen in performing analytical continuation
to let's say the Euclidean space.
\par
It is always possible to find (for a constructive approach
see \cite{Cola}) for any even-dimensional ($D=2n$)
$\Gamma$-structure a Weyl representation with the further property that
the $\Gamma^\mu$ matrices are all symmetric or antisymmetric under
transposition ($\Gamma^\mu = \pm {\Gamma^\mu}^T$) and moreover the
number of symmetric equal the number of antisymmetric $\Gamma^\mu$
matrices ($=n)$.\par In odd dimensional spacetimes an extra
symmetric matrix, the $\Gamma^5$ introduced in (\ref{12}) is
presents.\par A Wick rotation of a timelike $\overline{\mu}$
direction into a spacelike direction is represented on
$\Gamma$-matrices by the rescaling $\Gamma^{\overline{\mu}}\mapsto
i\Gamma^{\overline{\mu}}$, while the remaining $\Gamma$-matrices
are left unchanged. Clearly the symmetric or antisymmetric
character of the $\Gamma^{\overline{\mu}}$ matrix is not affected
by a Wick rotation.\par In a Weyl-represented even-dimensional
$\Gamma$-structure we can introduce two inequivalent charge
operators (i.e. realizing inequivalent $\{\Gamma^\mu, A, B,C\}$
assignments, see the discussion in the previous section) $C_S$ and
$C_A$ defined as follows
\begin{eqnarray}
C_S &=&\prod_{i_S= 1,...,n}\Gamma^{i_S}            \nonumber \\
C_A &=&\prod_{i_A=1,...,n}\Gamma^{i_A} \label{15}
\end{eqnarray}
the products being restricted to symmetric (and respectively
antisymmetric) $\Gamma$-matrices. As in the definition of the
matrix $A$ (\ref{8}), the ordering of the products is irrelevant.
Please notice that the index ($S$ or $A$) labeling $C$ reflects
the construction, via symmetric or antisymmetric matrices, of the
corresponding charge-conjugation operator and not its
(anti)-symmetry property which is expressed by formula (\ref{14}).
{}From (\ref{8}) and (\ref{15}) we obtain the relation $C_A\equiv
C_S\Gamma^5$.
\par
\par Clearly $C_S$ and $C_A$ are left invariant by Wick rotations
up to an arbitrary phase, implying the convenience of the Weyl
basis in discussing such an issue.
\par In odd-dimensional spacetimes a charge operator $C$ can be
introduced by using both formulas in (\ref{15}). Due to the
presence in this case of the extra $\Gamma^5$ among the symmetric
matrices, the two definitions indeed collapse into a single one
(modulo an arbitrary phase), recovering the well-known result that
there exists a unique $\{\Gamma^\mu, A, B, C \}$-assignment, up to
unitary transformations, in odd dimensions.\par We recall that the
$A$ matrix is defined in (\ref{8}), while $B_{S,A}$ are introduced
from (\ref{5}) as
\begin{eqnarray}
B_{S,A} &=& A\cdot{C_{S,A}}^T \label{16}
\end{eqnarray}
If we take into account the fact that timelike $\Gamma$-matrices
are hermitian, it is just a matter of tedious but straightforward
computations to check for both $(S,A)$-cases, which ($\pm 1
$)-signs correspond to $\eta_S$, $\eta_A$, as well as
$\varepsilon_S$, $\varepsilon_A$, introduced in the formulas
(\ref{4}) and (\ref{6}).
\par
In an ($s,t$) even-dimensional spacetime we obtain the following
table
\begin{center}
\begin{tabular}{|c|c|c|c|c|}
\hline
 $\spadesuit$& $0$ & $2$ &$ 4$ &$ 6$ \\ \hline
$\eta_S$ & $+$ &$ -$ & $+$ & $-$ \\ \hline $\eta_A $ & $-$ &$ +$
&$ -$ & $+$
\\ \hline $\varepsilon_S $ & $+$ &$ + $& $- $&$ -$ \\ \hline
$ \varepsilon_A $ &$ +$
&$ -$ & $- $& $+$
\\ \hline
\end{tabular}
\end{center}
\begin{eqnarray}
&&\label{17}
\end{eqnarray}
where the even values characterizing
the columns correspond to
\begin{eqnarray} X&=& s-t\quad mod\quad 8
\label{18}
\end{eqnarray}
A similar table can be produced for odd-dimensional spacetimes. In
this case no splitting between the $S,A$-cases is produced
\begin{center}
\begin{tabular}{|c|c|c|c|c|}
\hline
  $\spadesuit $ & $1$ & $3$ & $5$ & $7$ \\ \hline
  $\eta$ & $-$ & $+$ &$ -$ & $+$ \\ \hline
  $\varepsilon $& $+ $&$ -$ & $-$ & $+$ \\ \hline
\end{tabular}
\end{center}
\begin{eqnarray}
&&\label{19} \end{eqnarray} As above the columns are marked by $X$
given by (\ref{18}).\par Another sign, denoted by $\xi $ and
important for later considerations, is introduced through the
position
\begin{eqnarray}
B\Gamma^5 B^\dag &=& \xi \Gamma^5 \label{20}
\end{eqnarray}
where $\Gamma^5$ is the timelike extra-$\Gamma$ matrix given in
(\ref{12}). We obtain
\begin{center}
\begin{tabular}{|c|c|c|c|c|}
\hline
  $\spadesuit$ & $0$ & $2$ & $4$ & $6$ \\ \hline
  $\xi$ & $+$ & $-$ & $+$ & $-$ \\ \hline
\end{tabular}
\end{center}
\begin{eqnarray}
&&\label{21}\end{eqnarray}

Here as well columns correspond to $X$ given in (\ref{18}).\par
The Majorana reality condition on spinors is a constraint on the
charge-conjugated spinor $\psi^c$ introduced in (\ref{9}), imposed
to satisfy
\begin{eqnarray}
\psi^c &=& \psi \label{22}
\end{eqnarray}
Such a constraint can be consistently set only when $\varepsilon =
+1$ (due to the combined result of applying the complex
conjugation on (\ref{9}) and the formula (\ref{7})).\par One of
the consequences read from the table (\ref{17}) is the well-known
result that Majorana spinors do not exist in the euclidean
$4$-dimensional space.\footnote{Enlarged reality conditions which
are applicable when $\varepsilon=-1$, like the $SU(2)$-reality
condition proposed be Wetterich \cite{Wett}, will not be discussed
in the present paper.}\par The table (\ref{17}) is useful for
another purpose. It allows reading which choice of the $C_S$,
$C_A$ charge-conjugation operators should be adopted to mantain a
Majorana reality condition if a Wick analytical continuation is
performed. Indeed in the $0$-column both the $C_S$ and $C_A$
charge-conjugation operators are consistent with the Majorana
condition. Therefore e.g. the $(2,2)$ spacetime supports
inequivalent Majorana spinors, based either on $C_S$ or on $C_A$;
conversely for the standard $(3,1)$-Minkowskian spacetime the
Majorana condition is only defined w.r.t. $C_S$. Recalling the
property that $C_S$, $C_A$ are left unchanged by Wick rotation, it
turns out that only the $(2,2)$ $C_S$-Majorana spinors are Wick
related to the $(3,1)$ Majorana spinors.\par An euclidean space
which supports Majorana spinors is the $10$-dimensional one. We
obtain the two following chains of Wick-related Majorana spinors:
\begin{eqnarray}
C_S&:& (10,0)\rightarrow (9,1) \rightarrow (6,4)\rightarrow (5,5)
\rightarrow (2,8) \rightarrow (1,9)\nonumber\\ C_A&:& (0,10)
\rightarrow (1,9) \rightarrow (4,6)\rightarrow (5,5) \rightarrow
(8,2) \rightarrow (9,1) \label{23}
\end{eqnarray}
The three potentially ambiguous cases are $(9,1)$, $(5,5)$ and
$(1,9)$ which present both kinds of Majorana spinors.

\section{Inequivalent real Clifford-Weyl structures.}

Let the Clifford $\Gamma$-structure in the specific Weyl basis
given above be denoted a
Clifford-Weyl structure. In this section we provide an answer to
the question: how many inequivalent real Clifford-Weyl structures
do exist? To provide a solution we introduce an appropriate index
labeling inequivalent structures. \par The mathematical
formulation of the problem is better phrased as finding the
classes of equivalence of $\Gamma$-matrices up to orthogonal
conjugation
\begin{eqnarray}
\forall \mu , \quad \Gamma^\mu &\mapsto & O\Gamma^\mu
O^T\label{24}
\end{eqnarray}
with $O$ $2^{[{D\over 2}]}\times 2^{[{D\over 2}]}$ real-valued,
orthogonal ($OO^T=O^TO={\bf 1}$) matrices.\par We already
mentioned that the fundamental Pauli theorem guarantees that
$\Gamma$-matrices are uniquely represented up to unitary
conjugation; they however fit into different classes when just
orthogonality is concerned. \par One could ask whether this
well-posed mathematical problem has sensible physical
consequences. Indeed, as far as quantum mechanics is concerned,
equivalent descriptions are provided by unitary-transformed states
in a given Hilbert space. However, if some reality condition has
to be imposed, it may well restrict the class of allowed
transformations to be the orthogonal ones. Indeed this happens
when e.g. the Majorana reality condition is imposed on spinors.
Later we comment more on that.
\par
The above mathematical problem finds the following solution. \par
Let an index $I$ be defined for a $D$-dimensional
$(s,t)$-spacetime ($D=s+t$) through the position
\begin{eqnarray}
I&=& {1\over 2^{([{D\over 2}]+1)}}\cdot tr ( \Gamma^\mu
{\Gamma_\mu}^\ast ) \label{25}
\end{eqnarray}
The sum over repeated indices is understood. The normalization is
chosen for a matter of convenience and as before $[{D\over 2}]$
denotes the integral part of ${D\over 2}$.\par $I$ is clearly left
invariant by orthogonal transformations (\ref{24}) while it is
affected by unitary conjugations of $\Gamma$-matrices. It can be
therefore used to label inequivalent classes of $\Gamma$-matrices
up to orthogonal conjugation.\par In a Weyl basis $I$ can be
easily computed. Indeed, as previously recalled, in such a basis
$\Gamma$-matrices are either symmetric or antisymmetric.
${\Gamma^\mu}^\ast$ coincides with $\Gamma^\mu$ up to a sign which
is determined by both the time-like or space-like character of the
$\mu$ direction, as well as the (anti)-symmetry nature of
$\Gamma^\mu$. It is a matter of straightforward computations to
check the following results.\par $i)$ Let us consider at first an
even ($D=2n$) dimensional spacetime. We denote as $t_A$ ($s_A$)
the number of time-like (space-like) directions associated to
antisymmetric $\Gamma$-matrices. The number of symmetric timelike
(spacelike) matrices is therefore $t-t_A$ ($s-s_A$). In a Weyl
basis the equality $t_A + s_A = n$ holds. For a Weyl assignment
with $t_A$ antisymmetric timelike matrices the index $I$ takes the
value
\begin{eqnarray}
I &=& t-2t_A \label{26}
\end{eqnarray}
Let us introduce $m$ given by $m=\min (s,t)$. We are free to
choose among $m+1$ different Weyl assignments, $t_A=0,..., m$,
each one leading to a different value for $I$ and therefore
inequivalent under orthogonal conjugations. Indeed we obtain $m+1$
possible values for $I$,
\begin{eqnarray}
-m + 2j &,&\quad\quad\quad j=0,...,m\label{27}
\end{eqnarray}
in the even-dimensional case.\par $ii)$ Let $D=2n+1$ ($s+t=2n+1$)
be an odd-dimensional spacetime ($s+t=2n+1$). An extra (the
generalized-$\Gamma^5$ matrix here denoted $\Gamma^{2n+1}$)
symmetric matrix is present w.r.t. the previous case. It could be
associated either to a time-like or to a space-like direction
according to the sign
\begin{eqnarray}
{\Gamma^{2n+1}}^2 &=& \kappa = \pm 1\label{28}
\end{eqnarray}
Let as before $t_A$ be the number of antisymmetric timelike
$\Gamma$-matrices. The computation of the index $I$ follows the
same steps. We obtain
\begin{eqnarray}
I &=& t - 2t_A -{\textstyle{1\over 2}} (1-\kappa ) \label{29}
\end{eqnarray}
Notice that for a fixed $(s,t)$-spacetime the parity of $I$ is
determined by the sign of $\kappa$:
\begin{eqnarray}
(-1)^I &=& \kappa (-1)^t\label{30}
\end{eqnarray}
or conversely
\begin{eqnarray}
\kappa &=& (-1)^{t+I} \label{31}
\end{eqnarray}
which implies that the timelike or spacelike character of
$\Gamma^{2n+1}$ cannot be reverted by orthogonal conjugations. As
before let us put $m=\min (s,t)$.\par $\kappa$ can be arbitrary
chosen unless $m=0$. $I$ can assume $2m+1$ different values
labeling corresponding inequivalent classes. For $m\neq 0$ they
are given by
\begin{eqnarray}
&&m-2j -{\textstyle{1\over 2}} (1-\kappa )\nonumber\\ &&\kappa
=\pm 1, \quad\quad j = 0,1,..., m-{\textstyle{1\over 2}} (1+\kappa
) \label{32}
\end{eqnarray}
For $m=0$ either we have $I= +1$ in the $(2n+1,0)$ case or $I=-1$
in the $(0,2n+1)$ case.\par Let us discuss now a possible
application of the above construction to the Majorana reality
condition. A standard result (see \cite{Kugo}) states that for
$\varepsilon = 1$, i.e. the consistency requirement for the
Majorana condition, the $B$ matrix introduced in (\ref{5}) can be
unitary-transformed (\ref{10}) to the identity matrix
\begin{eqnarray}
&& \exists U \quad s.t. \quad U^\ast B U^\dag ={\bf 1} \label{33}
\end{eqnarray}
The choice $B\equiv {\bf 1}$ corresponds to the so-called Majorana
representation ($\psi^c =\psi^\ast$). The orthogonal
transformations are the unitary transformations acting on $B$ and
preserving the Majorana representation
\begin{eqnarray}
U^\ast {\bf 1} U^\dag ={\bf 1} &\Rightarrow & U U^T = U^T U ={\bf
1} \label{34}
\end{eqnarray}
From (\ref{4}) in the Majorana representation we have
${\Gamma^{\mu}}^\ast=\eta \Gamma^\mu $, so that the index $I$
takes the value
\begin{eqnarray}
I &=& \eta D \label{35}
\end{eqnarray}
In this particular case the information furnished by the index $I$
is reduced to the same information provided by $\eta,
\varepsilon$. The logics behind is however different.
$\eta,\varepsilon$ label inequivalent classes under unitary
transformations of a richer $\{\Gamma^\mu, A, B, C\}$-structure,
while $I$ corresponds to inequivalent classes of orthogonal
transformations of just a $\Gamma$-structure (in the Majorana
realization). It deserves a careful investigation to determine
whether for other choices of reality conditions which can select
physical fields (such as the $SU(2)$-Majorana condition on
spinors) the index $I$ can refine the standard classification and
be physically meaningful. In a different but related context we
already found \cite{Cola} a physical application where the index
plays a non-trivial role.

\section{Free Hermitian actions.}

The most general lagrangian involving free spinorial fields is
given by
\begin{eqnarray}
&&\alpha\cdot {\overline \psi}\Gamma^\mu\partial_\mu\psi +\beta
\cdot{\overline\psi}\psi +\gamma\cdot{\overline
\psi}\Gamma^\mu\Gamma^5\partial_\mu\psi +\delta\cdot
{\overline\psi}\Gamma^5\psi \label{36}
\end{eqnarray}
The third (pseudokinetic) and the fourth (pseudomassive) term
involve the $\Gamma^5$ matrix defined in (\ref{12}) and are
present in even $D$-dimensional spacetimes only.\par The
transposition acting on anticommuting fields $\zeta,\psi$
satisfies
\begin{eqnarray}
(\zeta\cdot\psi)^T&=& -\psi^T\cdot\zeta^T \label{37}
\end{eqnarray}
while the hermitian conjugation can be conventionally defined,
without losing generality, according to
\begin{eqnarray}
(\zeta\cdot\psi)^\dag &=& \psi^\dag\cdot\zeta^\dag \label{38}
\end{eqnarray}
(as for the complex conjugation, it follows from (\ref{37}) and
(\ref{38})).\par Demanding the hermiticity of the action, i.e. of
the (\ref{36}) lagrangian, fixes unambiguously the nature, real or
imaginary, of the coefficients in (\ref{36}). Straightforward
computations lead to the table
\begin{center}
\begin{tabular}{|c|c|c|c|c|}
  \hline
  $\spadesuit $ & $0$ & $1$& $2$ & $3$ \\ \hline
  $\alpha$ & ${\bf R }$& ${\bf I}$ & ${\bf I}$ & ${\bf R}$ \\
  \hline
  $\beta$ & ${\bf R}$ & ${\bf R}$  & ${\bf I}$ & ${\bf I}$ \\
  \hline
  $\gamma$& ${\bf I}$ & ${\bf I}$ & ${\bf R}$ & ${\bf R}$ \\
\hline  $\delta $& ${\bf R}$ & ${\bf I}$ & ${\bf I}$ & ${\bf R}$
\\ \hline
\end{tabular}
\end{center}
\begin{eqnarray}
&&\label{39}
\end{eqnarray}
the columns are labeled by $t\quad mod\quad 4,\quad$ $t$ being the
number of time-like dimensions.\par The table above is useful e.g.
in finding mass-shell properties. Indeed in the case of a theory
involving, let's say, massive and/or pseudomassive terms, the
mass-shell condition reads as follows
\begin{eqnarray}
p^2 &=& {{1\over \alpha^2}} (\delta^2-\beta^2)
\end{eqnarray}
where $\alpha,\beta,\delta$ enter (\ref{36}). The hermiticity
requirement (\ref{39}) allows to set e.g. in the ($3,1$) Minkowski
spacetime $\alpha = i$, $\beta = m$, $\delta = i m_5$, with
$m,m_5$ real, so that $p^2=m^2 + {m_5}^2$ is necessarily positive.
In a $(2,2)$-spacetime we only need to change the definition of
$\beta$, which must be imaginary, by setting $\beta = i m $. The
mass-shell condition reads $p^2 = {m_5}^2 - m^2$. A vanishing
value can be found even for $m, m_5 \neq 0$ provided that $m=m_5$.

\section{Majorana constraints on the dynamics.}

In this section we analyze the constraints put by the (\ref{9})
Majorana condition on the dynamics of free spinors.\par {} From
the (\ref{36}) lagrangian we derive the equation of motion
\begin{eqnarray}
\alpha\Gamma^\mu\partial_\mu\psi +\beta\psi
+\gamma\Gamma^\mu\Gamma^5\partial_\mu\psi +\delta\Gamma^5\psi &=&0
\label{41}
\end{eqnarray}
The above equation of motion is compatible with the (\ref{9})
Majorana condition provided the coefficients are constrained to
satisfy
\begin{eqnarray}
{\alpha^\ast} &=& \chi\cdot (\eta \alpha ) \nonumber\\
{\beta^\ast}&=& \chi\cdot\beta\nonumber\\ {\gamma^\ast} &=&
\chi\cdot (\eta\xi\gamma )\nonumber\\ \delta^\ast &=& \chi \cdot
(\xi \delta )
\end{eqnarray}
where $\eta$, $\xi$ have been introduced in (\ref{4}) and
(\ref{6}) respectively. The common factor $\chi$, as far as the
equation of motion alone is concerned, is an arbitrary phase
\begin{eqnarray}
\mid \chi \mid^2 &=& 1 \label{43}
\end{eqnarray}
The derivation of the (\ref{41}) equation of motion from a
lagrangian puts further constraints. Each Majorana-constrained
${\cal L}_i$ ($i=1,...,4$) term appearing in (\ref{36}), in order
to be non-vanishing, must be symmetric, i.e.
\begin{eqnarray}
{{\cal L}_i}^T &=& {\cal L}_i
\end{eqnarray}
For an $(s,t)$-spacetime ($s+t=D$) this so happens when the
following signs assume the $+1$ value:
\par
$i)$ for the kinetic term the sign is $\lambda$ given by
\begin{eqnarray}
\lambda &=& -\varepsilon\eta^{t+1}(-1)^{\textstyle{t\over
2}(t+1)}\label{45}
\end{eqnarray}

$ii)$ for the massive term, $\mu$
\begin{eqnarray}
\mu &=& -\varepsilon \eta^t (-1)^{\textstyle{t\over
2}(t-1)}\label{46}
\end{eqnarray}

$iii)$ for the pseudokinetic term, $\lambda_5$
\begin{eqnarray}
\lambda_5 &=& \lambda (-1)^{\textstyle{D\over 2}}
\end{eqnarray}

$iv)$ for the pseudomassive term, $\mu_5$
\begin{eqnarray}
\mu_5 &=& \mu (-1)^{\textstyle{D\over 2}}
\end{eqnarray}

The following tables, specifying which spacetimes support the
existence of non-vanishing kinetic and massive terms, can be
produced. For even-dimensional spacetimes we have\\
\begin{center}
\begin{tabular}{|c|c|c|c|c|}\hline
 $ \spadesuit$ & $0$ & $2$ & $4$ & $6$ \\ \hline
  $\lambda_S$ & $1,2$ & $0,1$ &  &  \\ \hline
  $\lambda_A$ & $0,1$ &  &  & $1,2$ \\ \hline
  $\mu_S$ & $2,3$ & $1,2$ &  &  \\ \hline
  $\mu_A$ & $1,2$ &  &  & $2,3$ \\ \hline
\end{tabular}
\end{center}
\begin{eqnarray}
&&\label{49}
\end{eqnarray}

Some comments are in order. The columns are labeled by $X=s-t\quad
mod\quad 8$. The index $S$ or $A$ is referred to the corresponding
charge-conjugation (either $C_S$ or $C_A$). The entries are
evaluated only  when $\varepsilon =1$ (Majorana consistency
requirement); the $\sharp$'s in the entries specify for which
number of $t$ time-like directions
\begin{eqnarray}
t&=&\sharp \quad mod \quad 4
\end{eqnarray}
the sign in the associated row assumes the $+1$ value.\par
Similarly, for odd-dimensional spacetimes, we have\\
\begin{center}
\begin{tabular}{|c|c|c|c|c|}
  \hline
  $\spadesuit$ & $1$ & $3$ & $5$ & $7$ \\ \hline
  $\lambda$ & 0,1 &  &  & $1,2$ \\ \hline
  $\mu$ & $1,2$ &  &  & $2,3$ \\ \hline
\end{tabular}
\begin{eqnarray}
&&\label{51}
\end{eqnarray}
\end{center}
(same meaning for the symbols).\par The next question to be
answered is whether the action associated to the (\ref{36})
lagrangian admits a charge-conjugation which allows to
consistently introduce the Majorana condition. This point has been
raised in \cite{DeAn}. The existence of a charge conjugation
requires
\begin{eqnarray}
{\cal L}^\ast &=& {\cal L}
\end{eqnarray}
and is automatically guaranteed from the non-vanishing condition
${\cal L}^T={\cal L}$ once assumed the hermiticity of the action
(${\cal L}^\dag = {\cal L}$), i.e. when the coefficients are
chosen to satisfy the table (\ref{39}).\par It turns out that the
phase $\chi$ appearing in (\ref{43}) is no longer arbitrary now
but fixed to be
\begin{eqnarray}
\chi &=& -\eta^t
\end{eqnarray}

{} From the (\ref{49}) and (\ref{51}) tables above we can extract
some particular results, e.g. that massive lagrangians for
Majorana spinors exists in\\ $i)$
 $t= 1 \quad mod\quad 4$ spacetimes (for $\eta=-1$)
 when\par
 $ia)$ $s-t = 0\quad mod \quad 8$ (for the $C_A$ charge-operator),\par
 $ib)$ $s-t = 2\quad mod\quad 8$ (for the $C_S$ charge-operator),\par
 $ic)$ $s-t = 1 \quad mod\quad 8 $,
 \\
 as well as in\\
 $ii)$ $t=2\quad mod\quad 4$ (for $\eta=+1$) when\par
 $iia)$ $s-t=0\quad mod\quad 8$ (for the $C_S$ charge-operator),\par
 $iib)$ $s-t = 6 \quad mod \quad 8 $ (for the $C_A$ charge-operator), \par
 $iic)$ $s-t=7 \quad mod\quad 8$.\par
 The role of $s,t$ can be interchanged as recalled in section
 $2$.\par
In the case of odd-dimensional spacetimes the table (\ref{51})
provides further information. Kinetic ($K$) or massive ($M$) terms
are only allowed in $D$-dimensional spacetimes according to
\begin{eqnarray}
&& D= 1\quad mod\quad 8 \quad \{K\}\nonumber\\ && D=3 \quad
mod\quad 8\quad \{K,M\}\nonumber\\ && D=5 \quad mod \quad 8\quad
\{M\}\nonumber\\ && D=7 \quad mod \quad 8 \quad \{...\}
\end{eqnarray}
Up to $D=11$ dimensions the list of odd-dimensional spacetimes
supporting Majorana spinors is given by
\begin{eqnarray}
\{K,M\}&:& (2,1), \quad (10,1), \quad (9,2), \quad(6,5)\nonumber\\
\{K\} &:&  (1,0),\quad (9,0),\quad (8,1),\quad (5,4)\nonumber\\
\{M\} &:& (2,3)\nonumber\\ \{...\} &:& (7,0),\quad (4,3)
\end{eqnarray}

For even-dimensional spacetimes (up to $D=10$) an useful table can
be written\\
\begin{center}\begin{tabular}{|c|c|c|c|c|c|}
\hline
  $\spadesuit$ & $2$  & $4$ & $6$  &  $8$  & $10$  \\ \hline
  $0$ & $S- K P$ & $$ & $A+$
   & $\begin{array}{l}
  S+ \\
  A-K
\end{array}$
& $S - K P$ \\  \hline
  $1$ & $\begin{array}{l}
 S + K P \\
  A - K M
\end{array}$& $S-KMP$ & $$ & $A+K$ & $\begin{array}{l}
 S+KP \\
  A-KM
\end{array}$ \\  \hline
  $2$ & $A+KM$ & $\begin{array}{l}
 S+KMP \\
  A-
\end{array}$ & $S-$ & $$ & $A+KM$ \\  \hline
  $3$ & $\bullet$ & $A+$ & $\begin{array}{l}
 S+ \\
  A-
\end{array}$ & $S-$ & $$ \\  \hline
  $4$ & $\bullet$ & $$ & $A+$ & $\begin{array}{l}
 S + \\
  A-K
\end{array}$ & $S-KP$ \\  \hline
  $5$ & $\bullet$ & $\bullet$ & $$ & $A+K$ & $\begin{array}{l}
 S+KP \\
  A-KM
\end{array}$\\  \hline
  $6$ & $\bullet$ & $\bullet$ & $S-$ & $$ & $A+KM$ \\  \hline
  $7$ & $\bullet$ & $\bullet$ & $\bullet$ & $A+$ & $$ \\  \hline
  $8$ & $\bullet$ & $\bullet$ & $\bullet$ & $\begin{array}{l}
 S + \\
  A-K
\end{array}$ & $S-KP$ \\  \hline
  $9$ & $\bullet$ & $\bullet$ & $\bullet$ & $\bullet$ &
  $\begin{array}{l}
 S +KP \\
  A-KM
\end{array}$ \\  \hline
  $10$& $\bullet$ &
  $\bullet$ & $\bullet$ & $\bullet$ & $A+KM$ \\ \hline
\end{tabular}
\end{center}
\begin{eqnarray}
&&
\end{eqnarray}
It contains the following informations. Columns are labeled by
$D$, rows by $t$. Each entry is evaluated for $\varepsilon =1$.
The presence of $S$ or $A$ denotes if the corresponding
charge-operator defines a Majorana spinor. The sign ($\pm$)
represents the corresponding value of $\eta$. The presence of
$K,M,P$ denotes if the kinetic ($K$), massive ($M$) or
pseudomassive ($P$) term in the (\ref{36}) lagrangian can be
non-vanishing. These last two terms have been evaluated only when
the corresponding kinetic term is nonzero. The pseudokinetic term
has not been inserted here since its physical interpretation is
problematic (due to the presence of negative-normed states, which
however can be eliminated if projected out, as in the
Majorana-Weyl case discussed in the next section).
\par
Since the same results are repeated $mod\quad 8$, both in $s$ and
in $t$, the following compact information can be extracted.
Majorana spacetimes with a non-vanishing kinetic term can be found
in
\begin{eqnarray}
D&=& 0\quad mod \quad 8: \quad\{ A, K \}\nonumber\\ D&=&2\quad mod
\quad 8: \quad either\quad \{S, KP\}\quad or \quad \{A,
KM\}\nonumber\\ D&=& 4 \quad mod\quad 8: \quad \{S, KMP\}
\end{eqnarray}
In $D=2, 10,...$ either a massive or a pseudomassive term could be
present, according to the choice of the charge-conjugation
operator. Simultaneous presence of massive and pseudomassive terms
is allowed in $D=4,12,...$ dimensions only.

\section{The Majorana-Weyl conditions.}

In order to make this paper self-consistent we review in this
section the status of Majorana-Weyl spinors and present a complete
list of results.\par In $D=2n$ even-dimensional spacetimes the
projectors
\begin{eqnarray}
P_{R,L} &\equiv & (\frac{{\bf 1} \pm \Gamma^5}{2})
\end{eqnarray}
(where $\Gamma^5$ has been introduced in (\ref{12})) allow
defining chiral (Weyl) spinors $\psi_{R,L}$ as
\begin{eqnarray}
\psi_{R,L} &=& P_{R,L}\psi \label{59}
\end{eqnarray}
Majorana-Weyl spinors, satisfying both the condition (\ref{9}) and
the projection (\ref{59}), can be consistently defined (see
\cite{Kugo}) in spacetimes such that
\begin{eqnarray}
s-t &=& 0\quad mod\quad 8 \label{60}
\end{eqnarray}
therefore in particular in all $(n,n)$ spacetimes. Up to $10$
dimensions the remaining spacetimes supporting Majorana-Weyl
spinors are the euclidean $(8,0)$ space and the minkowskian
$(9,1)$ spacetime.\par In any spacetime satisfying (\ref{60})
Majorana-Weyl spinors can be introduced for both $C_S$ and $C_A$
charge-conjugation operators. The list of results presented below
holds in both cases.\par Let us first recall that ${\overline\psi}
=\psi^T C$ under the condition (\ref{9}) and that moreover the
$C_{S,A}$ charge-operator is respectively block-diagonal or
block-antidiagonal according if $n$ is even or odd. As a
consequence kinetic ($K$) and massive ($M)$ terms can either mix
(denoted in such case as $K_{xy}, M_{xy}$) chiralities or not
($K_{xx}, M_{xx})$. We can write
\begin{eqnarray} K_{xx} &\equiv &
{\psi_{R,L}}^T C \Gamma^\mu \partial_\mu \psi_{R,L}\nonumber\\
M_{xx} &\equiv & {\psi_{R,L}}^T C\psi_{R,L} \label{61}
\end{eqnarray}
and
\begin{eqnarray}
K_{xy} &\equiv & {\psi_R}^T C\Gamma^\mu \partial_\mu \psi_L +
\lambda {\psi_L}^T C\Gamma^\mu\partial_\mu \psi_R\nonumber\\
M_{xy} &\equiv & {\psi_R}^T C\psi_L +\mu {\psi_L}^T C\psi_R
\label{62}
\end{eqnarray}
The ``mixed" terms $K_{xy}, M_{xy}$ can always be chosen to be
non-vanishing. It is sufficient for this purpose to conveniently
fix the relative sign between the two terms in the r.h.s. of
(\ref{62}). This is done in (\ref{62}), the two signs $\lambda$
and $\mu$ coincide with their values given in (\ref{45}) and
(\ref{46}) respectively.\\ Conversely the $K_{xx}$ and $M_{xx}$
terms could be identically zero according to the (anti-)symmetry
properties of $C$.
\par
Let us now introduce $\upsilon= (-1)^n$. From the previous remarks
on the block-character of $C$ we have that
\begin{eqnarray}
K_{\upsilon = +1} \equiv K_{xy} \quad &,& \quad M_{\upsilon =+1}
\equiv M_{xx}\nonumber\\ K_{\upsilon=-1} \equiv K_{xx} \quad &,&
\quad M_{\upsilon = -1} \equiv M_{xy} \label{63}
\end{eqnarray}
The most general free lagrangian for Majorana-Weyl spinors in
$D=2n$ dimensions can be expressed as \begin{eqnarray} {\cal L}
&=& \alpha K_{\upsilon} +\beta M_{\upsilon} \end{eqnarray} The
formula (\ref{63}) specifies which kind of kinetic and which kind
of massive term could appear in $D=2n$. The coefficients
$\alpha,\beta$ are either real or imaginary according to the table
(\ref{39}).\par The last feature to be computed is in which
dimensions the $K_{xx}$, $M_{xx}$ terms are not identically
vanishing. The final results can be summarized in the following
table, which presents the types of allowed kinetic and massive
terms in accordance with the dimensionality $D$ of the spacetime
\begin{eqnarray}
&& D = 0\quad mod\quad 8,\quad \{ K_{xy}\}\nonumber\\ && D=2\quad
mod\quad 8, \quad \{ K_{xx}, M_{xy} \}\nonumber\\ && D=4\quad mod
\quad 8, \quad \{ K_{xy}, M_{xx}\}\nonumber\\ && D=6 \quad
mod\quad 8, \quad \{ M_{xy} \}
\end{eqnarray}
The list of results presented in this section removes any possible
ambiguities and completely determines all features of the free
actions for Majorana-Weyl spinors in any space-time.

\section{Conclusions.}

This paper has been devoted to discuss real structures in Clifford
algebras and Majorana conditions in any space-time. Without loss of
generality a specific Weyl
representation, useful for dealing with Wick rotations,
of Clifford algebras has been employed to analyze
$\Gamma$-structures and Majorana spinors. An index, which to our
knowledge has not been discussed before at least in the
physicists' literature, has been introduced. It classifies
$\Gamma$-structures up to orthogonal conjugations. \par For what
concerns Majorana spinors, some of the issues here discussed have
not been considered in previous papers. We can mention e.g. the
interplay between the hermiticity condition, the
charge-conjugation and the non-vanishing condition for the
(\ref{36}) lagrangian. The different role played by the
even-dimensional charge operators $C_S$, $C_A$ (\ref{15}),
invariant under Wick rotations, is another example.\par We have
furnished a series of tables presenting an exhaustive list of
results concerning Majorana and Majorana-Weyl spinors. They
include in particular the non-vanishing conditions in any given
space-time for kinetic, massive and pseudomassive terms, in
association with each charge operator $C_S$, $C_A$ (in even
dimensions), as well as the type of coefficients (\ref{39}) and
the kind of terms (in the Majorana-Weyl case) entering the free
lagrangian (\ref{36}).\par One of our main motivations for
presenting here such a systematic list of results concerns their
relevance in analyzing supersymmetries in generic pseudoeuclidean
spacetimes. Their connection with supergravities, strings, brane
dynamics, etc., could be explored (for a recent review of this
topic in standard Minkowskian spacetimes see e.g. \cite{West}). In
the introduction we mentioned why this issue could be important.
Problems like Kaluza-Klein compactifications, dimensional
reductions, analytical continuations to the euclidean spaces, are
among those which have to be addressed.

\vskip1cm \noindent{\Large{\bf Acknowledgments}} \\ {\quad}\\ We
are pleased to acknowledge J. A. Helay\"{e}l-Neto and L.P. Colatto
for both encouragement and helpful discussions. We are grateful to
C. Preitschopf for having pointed out an ambiguous statement made
in the preliminary version, which is now fully clarified.
 We are grateful to DCP-CBPF for the kind hospitality.

\end{document}